\documentclass{webofc}
\usepackage[varg]{txfonts}   
\begin{document}
\title{Observing with NIKA2Pol from the IRAM 30m telescope.}
%
%
\subtitle{Early results on the commissioning phase.}

\author{\firstname{A.} \lastname{Ritacco} \inst{\ref{IAS},\ref{IRAME}}\fnsep\thanks{\email{alessia.ritacco@ias.u-psud.fr}}
\and \firstname{R.} \lastname{Adam} \inst{\ref{LLR},\ref{CEFCA}}
\and  \firstname{P.} \lastname{Ade} \inst{\ref{Cardiff}}
\and  \firstname{H.} \lastname{Ajeddig} \inst{\ref{CEA1}}
\and  \firstname{P.} \lastname{Andr\'e} \inst{\ref{CEA1}}
\and  \firstname{A.} \lastname{Andrianasolo} \inst{\ref{IPAG}}
\and  \firstname{H.} \lastname{Aussel} \inst{\ref{CEA1}}
\and  \firstname{A.} \lastname{Beelen} \inst{\ref{IAS}}
\and  \firstname{A.} \lastname{Beno\^it} \inst{\ref{Neel}}
\and  \firstname{A.} \lastname{Bideaud} \inst{\ref{Neel}}
\and  \firstname{O.} \lastname{Bourrion} \inst{\ref{LPSC}}
\and  \firstname{M.} \lastname{Calvo} \inst{\ref{Neel}}
\and  \firstname{A.} \lastname{Catalano} \inst{\ref{LPSC}}
\and  \firstname{B.} \lastname{Comis} \inst{\ref{LPSC}}
\and  \firstname{M.} \lastname{De~Petris} \inst{\ref{Roma}}
\and  \firstname{F.-X.} \lastname{D\'esert} \inst{\ref{IPAG}}
\and  \firstname{S.} \lastname{Doyle} \inst{\ref{Cardiff}}
\and  \firstname{E.~F.~C.} \lastname{Driessen} \inst{\ref{IRAMF}}
\and  \firstname{A.} \lastname{Gomez} \inst{\ref{CAB}}
\and  \firstname{J.} \lastname{Goupy} \inst{\ref{Neel}}
\and  \firstname{F.} \lastname{K\'eruzor\'e} \inst{\ref{LPSC}}
\and  \firstname{C.} \lastname{Kramer} \inst{\ref{IRAME}}
\and  \firstname{B.} \lastname{Ladjelate} \inst{\ref{IRAME}}
\and  \firstname{G.} \lastname{Lagache} \inst{\ref{LAM}}
\and  \firstname{S.} \lastname{Leclercq} \inst{\ref{IRAMF}}
\and  \firstname{J.-F.} \lastname{Lestrade} \inst{\ref{LERMA}}
\and  \firstname{J.F.} \lastname{Mac\'ias-P\'erez} \inst{\ref{LPSC}}
\and  \firstname{P.} \lastname{Mauskopf} \inst{\ref{Cardiff},\ref{Arizona}}
\and \firstname{A.} \lastname{Maury} \inst{\ref{CEA1}}
\and \firstname{F.} \lastname{Mayet} \inst{\ref{LPSC}}
\and  \firstname{A.} \lastname{Monfardini} \inst{\ref{Neel}}
\and  \firstname{L.} \lastname{Perotto} \inst{\ref{LPSC}}
\and  \firstname{G.} \lastname{Pisano} \inst{\ref{Cardiff}}
\and  \firstname{N.} \lastname{Ponthieu} \inst{\ref{IPAG}}
\and  \firstname{V.} \lastname{Rev\'eret} \inst{\ref{CEA1}}
\and  \firstname{C.} \lastname{Romero} \inst{\ref{IRAMF}}
\and  \firstname{H.} \lastname{Roussel} \inst{\ref{IAP}}
\and  \firstname{F.} \lastname{Ruppin} \inst{\ref{MIT}}
\and  \firstname{K.} \lastname{Schuster} \inst{\ref{IRAMF}}
\and \firstname{Y.} \lastname{Shimajiri} \inst{\ref{CEA1},\ref{Kagoshima}, \ref{NAOJ}}
\and  \firstname{S.} \lastname{Shu} \inst{\ref{IRAMF}}
\and  \firstname{A.} \lastname{Sievers} \inst{\ref{IRAME}}
\and  \firstname{C.} \lastname{Tucker} \inst{\ref{Cardiff}}
\and  \firstname{R.} \lastname{Zylka} \inst{\ref{IRAMF}}}

\institute{\label{LPSC} Univ. Grenoble Alpes, CNRS, Grenoble INP, LPSC-IN2P3, 53, avenue des Martyrs, 3000 Grenoble, France
\and \label{LLR} LLR (Laboratoire Leprince-Ringuet), CNRS, \'Ecole Polytechnique, Institut Polytechnique de Paris, Palaiseau, France  
\and \label{CEFCA} Centro de Estudios de F\'isica del Cosmos de Arag\'on (CEFCA), Plaza San Juan, 1, planta 2, E-44001, Teruel, Spain 
\and \label{Cardiff} Astronomy Instrumentation Group, University of Cardiff, UK          
\and \label{CEA1} AIM, CEA, CNRS, Universit\'e Paris-Saclay, Universit\'e Paris Diderot, Sorbonne Paris Cit\'e, 91191 Gif-sur-Yvette, France     
\and \label{IPAG} Univ. Grenoble Alpes, CNRS, IPAG, 38000 Grenoble, France     
\and \label{IAS} Institut d'Astrophysique Spatiale (IAS), CNRS and Universit\'e Paris Sud, Orsay, France    
\and \label{Neel} Institut N\'eel, CNRS and Universit\'e Grenoble Alpes, France
\and \label{Roma} Dipartimento di Fisica, Sapienza Universit\`a di Roma, Piazzale Aldo Moro 5, I-00185 Roma, Italy       
\and \label{IRAMF} Institut de RadioAstronomie Millim\'etrique (IRAM), Grenoble, France 
\and \label{CAB} Centro de Astrobiolog\'ia (CSIC-INTA), Torrej\'on de Ardoz, 28850 Madrid, Spain
\and \label{IRAME} Instituto de Radioastronom\'ia Milim\'etrica (IRAM), Granada, Spain 
\and \label{LAM} Aix Marseille Univ, CNRS, CNES, LAM (Laboratoire d'Astrophysique de Marseille), Marseille, France
\and \label{LERMA} LERMA, Observatoire de Paris, PSL Research University, CNRS, Sorbonne Universit\'es, UPMC Univ. Paris 06, 75014 Paris,
France
\and \label{Arizona} School of Earth and Space Exploration and Department of Physics, Arizona State University, Tempe, AZ 85287         
\and \label{IAP} Institut d'Astrophysique de Paris, CNRS (UMR7095), 98 bis boulevard Arago, 75014 Paris, France
\and \label{MIT} Kavli Institute for Astrophysics and Space Research, Massachusetts Institute of Technology, Cambridge, MA 02139, USA 
\and \label{Kagoshima} Department of Physics and Astronomy, Graduate School of Science and Engineering, Kagoshima University, 1-21-35 Korimoto, Kagoshima, Kagoshima 890-0065, Japan
\and \label{NAOJ} National Astronomical Observatory of Japan, Osawa 2-21-1, Mitaka, Tokyo 181-8588, Japan
          }

\newcommand{\nika}{{\it NIKA}}
\newcommand{\nikad}{{\it NIKA2}}

\abstract{The NIKA2 polarization channel at 260 GHz (1.15 mm) has been proposed primarily to observe galactic star-forming regions and probe the critical scales between 0.01-0.05 pc at which magnetic field lines may channel the matter of interstellar filaments into growing dense cores. The NIKA2 polarimeter consists of a room temperature continuously rotating multi-mesh HWP and a cold polarizer that separates the two orthogonal polarizations onto two 260 GHz KIDs arrays. 
We describe in this paper the preliminary results obtained during the most recent commissioning campaign performed in December 2018.
We concentrate here on the analysis of the extended sources, while 
the observation of compact sources is presented in a companion paper \cite{hamza2019}.
We present preliminary NIKA2 polarization maps of the Crab nebula. We find that the integrated polarization intensity flux measured by NIKA2 is consistent with expectations. In terms of polarization angle, we are still limited by systematic uncertainties that will be further investigated in the forthcoming commissioning campaigns.}
\maketitle
\section{Introduction}
\label{intro}
The interstellar medium is permeated by large-scale magnetic fields \cite{planckxxxiii}. These magnetic fields are observed in  molecular clouds and are suggested to have a crucial role in  regulating the fragmentation of dense filaments and in channeling filament material into prestellar cores \cite{mckee,crutcher2012}. Observations of linearly-polarized continuum emission from magnetically-aligned dust grains at mm and submm wavelengths are a powerful tool to measure the morphology and structure of the magnetic field lines in star-forming clouds and dense cores \cite{crutcher2012}. 
The dust polarization maps provided by the {\it Planck} satellite have revealed a large-scale regular
morphology for the Galactic magnetic field (GMF). The orientation of this magnetic field tends to be parallel to low matter density and perpendicular to high matter density structures \cite{Planck2016}. Using complementary total intensity high angular resolution observations provided by the {\it Herschel} satellite \cite{andre2014,palmeirim2013,Arzouminian2011,Arzouminian2019} it has been shown that galactic filamentary structures are associated with an organized magnetic field topology at scales larger than 0.5 pc.
To probe the role played by the magnetic fields in the star formation mechanism, we need to explore smaller angular scales within the filaments. This could be performed with the polarization channel of NIKA2 \cite{NIKA2-Adam} at 260 GHz (1.15 mm), which will provide an in-depth view of the magnetic fields at the critical scales of 0.01 to 0.1 pc, thanks to an angular resolution of 11.2 arcsec. 
In addition, the NIKA2 camera has dual band capability,  providing simultaneous observations in total intensity at 150 GHz (2.05 mm) with an angular resolution of 17.5 arcsec. 

\section{NIKA2 polarization system and detection strategy}
\label{nika2polsys}
The NIKA2 polarization system consists of a room temperature continuously rotating half-wave plate (HWP) and a fixed polarizer mounted inside the cryostat at base temperature of ~150 mK \cite{NIKA2-Adam}.
The cold polarizer separates with high purity the two polarization onto two different KID arrays (see Figure~1).
Figure~2 shows the configurations in total power observations and polarization mode. In the latter, the HWP is placed in front of the cryostat window. The rotation of the HWP modulates the incident polarization. 
As a consequence, the polarization signal is shifted at the fourth harmonic of the HWP rotation frequency $\omega$ and can be recovered by using a demodulation technique, which consists of a numerical lock-in around the 4$\omega$ \cite{ritacco2017} frequency. This detection strategy allows a quasi-simultaneous observation of the three Stokes parameters $I$, $Q$, and $U$ although HWP-induced systematic effects are observed and need to be subtracted. The most important one corresponds to a parasitic signal at all harmonics of the HWP rotation frequency $\omega$. This systematic effect, which is common to all the pixels of the KID array, is called HWP Synchronous Signal (HWPSS) and it is corrected in the data analysis \cite{ritacco2017}.
Furthermore, we also observe in the NIKA2 polarization data intensity to polarization leakage, which constitutes to date the most critical systematic effect. From observations of Uranus, which can be considered as unpolarized, we have estimated a
maximum fractional instrumental polarization varying from 0.45~\% to 1.84~\% in Stokes $U$ and from 1.27~\% to 2.16~\% 
in Stokes $Q$ \cite{hamza2019}. A similar effect was observed for the NIKA camera and we developed a procedure to correct for it \cite{ritacco2017}. Although the intensity to polarization leakage in NIKA2 is about a factor 2 lower than in NIKA, it can not be corrected for using the NIKA procedure because it varies more significantly with elevation. Work is in progress to account for this effect and it is the priority of the forthcoming commissioning campaigns. In this paper, we concentrate on extended sources for which the intensity to polarization leakage effect is reduced when averaging across the source (see ~\cite{ritacco2017,ritacco2018}).


\begin{figure}
  \begin{center}
    \includegraphics[scale=0.3]{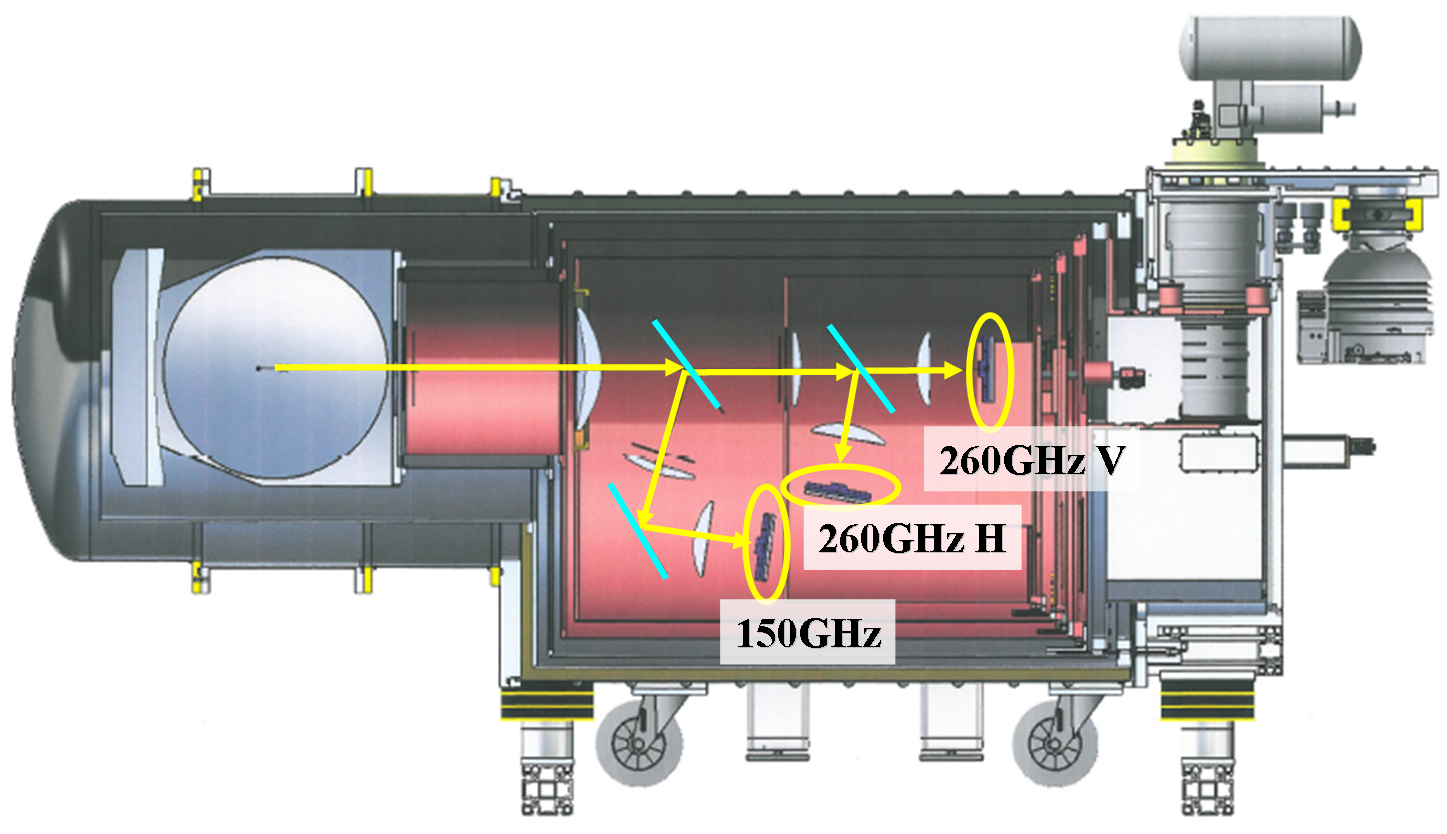}    
    \label{fig_cryo}       
    \caption{Cryostat scheme showing the two NIKA2 channels at 150 GHz (2.05 mm) and 260 GHz (1.15 mm). The two 1mm arrays recover the two orientations of the linear polarization.}
 \end{center}
\end{figure}
\begin{figure}
\begin{center}
\includegraphics[scale=0.5]{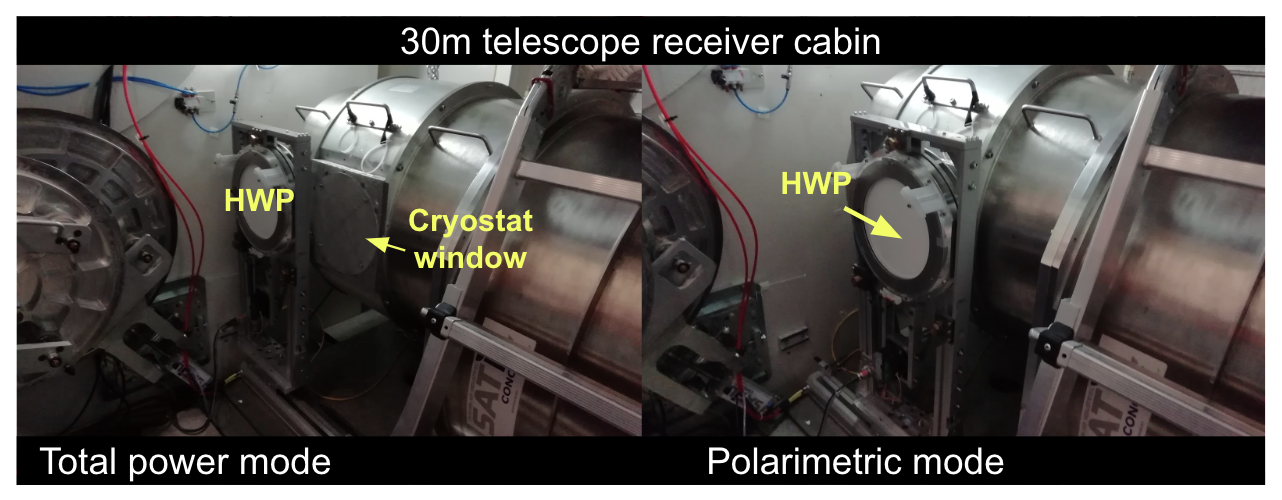}
 \label{fig_cabin}       
    \caption{The two images show the NIKA2 cryostat inside the receiver cabin of the IRAM 30m telescope. The HWP is placed in front of the cryostat entrance when we run observations in polarization mode.}
 \end{center}
\end{figure}

\section{NIKA2 first polarization light on the Crab Nebula}

For the NIKA2 commissioning, we concentrated on observations of the Crab Nebula, which were taken in December 2018.
The Crab nebula is a supernova remnant that is generally considered as a good polarization angle calibrator for polarization experiments (\cite{ritacco2018,aumont2019}). The spectral energy distribution in both intensity and polarization of the Crab nebula is well described by a single power law spectrum, as expected from synchrotron emission powered by a single population of relativistic electrons \cite{ritacco2018}. As a consequence, we expect the Crab nebula degree and polarization angle to be constant at radio and millimeter wavelengths.  This has been shown by \cite{ritacco2018} using intensity and polarization observations of the Crab nebula from 23 to 217 GHz. However, other emission contributions are expected and they could be investigated using NIKA2 among other experiments. \\

Figure~3 shows the 1.15 mm NIKA2 maps of Stokes $I$, $Q$, and $U$.  Polarized vectors, where the polarization intensity SNR is larger than 3$\sigma$, are also overplotted on the intensity map. 
The polarization intensity flux integrated over the extension of the Crab nebula is $12.3 \pm 0.1$ Jy (statistical uncertainties only). This value is consistent with the expectation from the polarization intensity SED given in \cite{ritacco2018}. For the Stokes $I$ map we obtain a total flux of 158.04$\pm$0.18 Jy (statistical uncertainties only). The discrepancy between the recovered flux in total intensity and the expected one \cite{ritacco2018} may be due to filtering effects that were not fully taken into account on this preliminary result. In terms of polarization angle, we find (-86.9$\pm$0.1)$^{\circ}$ (statistical uncertainties only), which is only marginally consistent with the averaged value of  (-87.7$\pm$0.3)$^{\circ}$  estimated at these frequencies by \cite{ritacco2018}. Although NIKA2 preliminary results are fairly consistent with Planck satellite results given in \cite{ritacco2018}, there are still few systematic effects observed on compact sources that need to be addressed.
Systematic uncertainties due to the HWPSS and to the intensity to polarization leakage are under investigation and will require extra commissioning campaigns, which are planned for early 2020.
At present we still observe significant discrepancy between the polarization angle calibration deduced from compact and extended sources, which is mainly related to the intensity to polarization leakage correction \cite{hamza2019}.

\begin{figure}
\begin{center}
\includegraphics[scale=0.45]{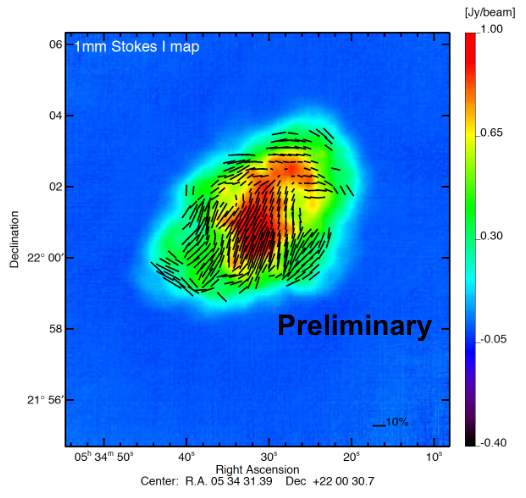}
\includegraphics[scale=0.45]{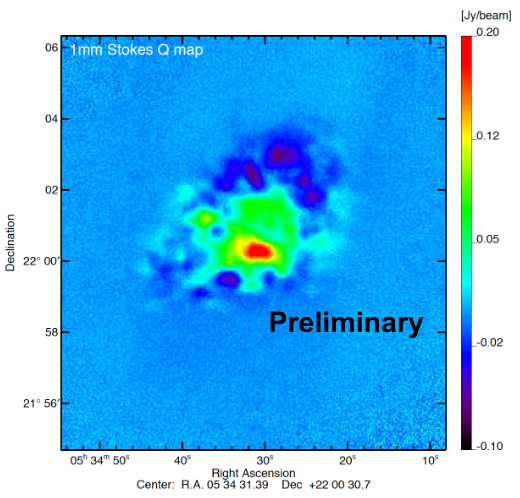}
\includegraphics[scale=0.45]{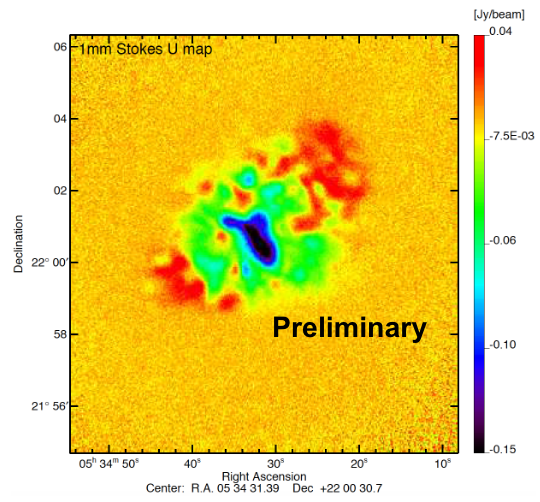}
\label{fig2}
\caption{NIKA2 Stokes parameter maps of the Crab Nebula at 1.15 mm. From left to right: intensity map with polarization vectors overplotted where the polarization intensity SNR > 3$\sigma$, Stokes Q map, and Stokes U map.}
\end{center}
\end{figure}

\section{Summary and conclusions}
The NIKA2 polarization channel at 260 GHz entered its commissioning phase in Autumn 2017, when the NIKA2 camera was officially offered to the community for total intensity observations. Because of hardware problems and bad weather, first reliable data were only obtained in December 2018.
During this commissioning campaign, we were able to observe well known compact (see \cite{hamza2019}) and extended sources including the Crab nebula. In this paper, we have presented the preliminary NIKA2 polarization maps of the Crab nebula. These maps allowed us to compute the integrated polarization flux, which is consistent with expectations from the polarization SED shown in~\cite{ritacco2018}. Preliminary NIKA2 measurements of the Crab polarization angle indicate fair agreement with expectations although systematic uncertainties are not fully understood to date.
Overall the current analysis is still limited by systematic effects and requires further investigation. Extra commissioning campaigns are planned for early 2020.


\section*{Acknowledgements}
We would like to thank the IRAM staff for their support during the campaigns. The NIKA dilution cryostat has been designed and built at the Institut Néel. In particular, we acknowledge the crucial contribution of the Cryogenics Group, and in particular Gregory Garde, Henri Rodenas, Jean Paul Leggeri, Philippe Camus. This work has been partially funded by the Foundation Nanoscience Grenoble and the LabEx FOCUS ANR-11-LABX-0013.  This work is supported by the French National Research Agency under the contracts "MKIDS", "NIKA"and ANR-15-CE31-0017 and in the framework of the "Investissements d’avenir” program (ANR-15-IDEX-02).  This work has benefited from the support of the European Research Council Advanced Grant ORISTARS under the European Union’s Seventh Framework Programme (Grant Agreement no.  291294).  F.R. acknowledges financial supports provided by NASA through SAO Award Number SV2-82023 issued by the Chandra X-Ray Observatory Center, which is operated by the Smithsonian Astrophysical Observatory for and on behalf of NASA under contract NAS8-03060.

%

\begin{thebibliography}{}
%
%
\bibitem{planckxxxiii}
Planck Collaboration Int. XXXIII. 2016, Astron.\ Astrophys.\ {\bf 586}, A13
\bibitem{mckee}
McKee, C. F. \& Ostriker, E. C., 
ARA\&A, {\bf 45}, 565 (2007)
\bibitem{crutcher2012}
Richard M.~Crutcher,
Annual Review of Astronomy and Astrophysics, vol. 50, p.29-63 (2012)
\bibitem{Planck2016}
Planck collaboration {\it et al.},
Astron.\ Astrophys.\ \textbf{594}, A1 (2016)

\bibitem{andre2014}
Ph. André {\it et al.}, 
in \textit{Protostars and Planets VI}, 27--51 (2014) 

\bibitem{palmeirim2013}
  P.~Palmeirim {\it et al.},
  Astron.\ Astrophys.\  {\bf 550}, A38 (2013)
  
\bibitem{Arzouminian2011}
D. Arzoumanian {\it et al.},
Astron.\ Astrophys.\ \textbf{529}, L6 (2011) 

\bibitem{Arzouminian2019}
D. Arzoumanian {\it et al.},
Astron.\ Astrophys.\ \textbf{621}, A42 (2019) 
\bibitem{NIKA2-Adam}
R.~Adam {\it et al.},
  Astron.\ Astrophys.\  {\bf 609}, A115 (2018)
\bibitem{ritacco2017}
  A.~Ritacco {\it et al.},
  Astron.\ Astrophys.\  {\bf 599}, A34 (2017)

  \bibitem{ritacco2018}
  A.~Ritacco {\it et al.},
  Astron.\ Astrophys.\  {\bf 616}, A35 (2018)

\bibitem{hamza2019}
H.~Ajeddig {\it et al.},
NIKA2 conference proceeding, 2019 

\bibitem{aumont2019}
J.~Aumont {\it et al.},
arXiv:1805.10475, submitted to Astron.\ Astrophys.\ 2019
%


%






%
%








  

  




\end{thebibliography}
%
%

\end{document}